# Enabling atomic resolution in convergent beam EMCD measurements by the use of patterned apertures

Hasan Ali[1], Devendra Negi[2,†], Tobias Warnatz[2], Björgvin Hjörvarsson[2], Jan Rusz[2], Klaus Leifer[1,*]

[1] *Electron Microscopy and Nano-Engineering, Applied Materials Science, Department of Engineering Sciences, Uppsala University, Box. 534, 751 21 Uppsala, Sweden*
[2] *Department of Physics and Astronomy, Uppsala University, Box 516, 751 20 Uppsala, Sweden*
[†] *Present address: Stuttgart Center for Electron Microscopy, Max Planck Institute for Solid State Research, Heisenbergstr.1, 70569 Stuttgart, Germany*

We give an experimental demonstration of two types of recently proposed ventilator apertures which can be used to acquire electron magnetic circular dichroic (EMCD) signals in zone axis orientation with high spatial resolution. To simplify the experimental procedures, we propose a third type of aperture and experimentally demonstrate the use of this modified ventilator aperture for the case of multiple symmetries in the diffraction patterns. To show the feasibility of the atomic resolution EMCD, EMCD signals are acquired for a range of beam convergence angles. High quality EMCD signals with convergence angles corresponding to atomic resolution electron probes are obtained.

If one were to build an instrument for investigating spatial dependence of quantum effects at interfaces, one of the determining design criteria would be the obtained spatial resolution. The required resolution is in the range of few atomic distances to few tens of nanometer [1, 2] which is close to ideal for transmission electron microscope (TEM). The TEM with its analysis volume of the specimen between a few $nm^3$ and a few $10 \, nm^3$ has been used for investigating quantum wires, quantum dots and interfaces incorporated in a thin film or a bulk material. In the analysis of nano-magnetic materials, TEM has played a very important role since it had enabled the analysis of such materials with a resolution of below 10 nm using techniques of electron holography and Lorentz microscopy [3-6]. In such analysis, the resolution of the TEM has bridged the gap between the atom by atom surface analysis in the STM [7] and the analysis using non-microscopy techniques like XMCD [8]. With recent developments in the technique of differential phase contrast (DPC) [9, 10] and electron magnetic circular dichroism (EMCD) [11-16], the perspective of achieving atomic resolution for the analysis of magnetic properties appears reachable. Whereas in current atomically resolved DPC work, the TEM lamella thickness is of a few nanometers, in the case of EMCD, typically sample thicknesses range from 10-50 nm making it thus possible to have the full quantum object in the analysis volume and addressing thus magnetic quantum phenomena. In classical EMCD experiments, two EELS spectra are acquired at two specific scattering angles in the diffraction plane and the difference of these spectra, called the EMCD signal, is used to determine the magnetic moments of the material. One of the difficulties of the EMCD technique operating at sub-nanometer to atomic resolution has been hitherto that high convergence angles of the focused electron probe are needed which have the drawback that the electron beam diffraction disks start to overlap in the diffraction plane making thus the q-selection needed to obtain the EMCD signal difficult. Another limitation to obtain atomic resolution EMCD is that most work hitherto was carried out having atomic planes parallel to the electron beam to reach the required diffraction geometry of a 2-beam or 3-beam condition [11] losing atomic column resolution. There is though one previous EMCD work carried out on a zone-axis orientation where a parallel electron beam of 50 nm in diameter was used in the experiments [17]. Nevertheless, Negi et al. [18] have shown in their simulations that an EMCD signal can be obtained from highly convergent electron beams and on the zone axis. In this paper, the use of symmetric multihole apertures, called in the following ventilator apertures, was proposed to obtain the EMCD signal. Here, we have fabricated such apertures and show, that from an optimization of those apertures EELS spectra can be acquired that contain a clear EMCD signal at electron probe sizes below 2Å enabling thus atomic resolution EMCD analysis.

In this Letter we first demonstrate EMCD signals that are obtained by the use of an 8-fold ventilator apertures at a beam convergence semi-angle of $\alpha = 5$ mrad in a high symmetry zone axis. Then, a modified ventilator aperture, enabling the simultaneous acquisition of both conjugated EELS spectra is fabricated and equally yields a clear EMCD signal. We then propose and experimentally demonstrate a simplified version of the ventilator aperture which has the freedom to be used in multiple crystal symmetries keeping a zone axis orientation. We also demonstrate clear EMCD signal with $\alpha = 10$ mrad which results in an electron probe size of 1.2 Å. The upper experimental limit of $\alpha$ is set by the spherical aberration of the instrument and an aberration corrected instrument would allow for even higher convergence angles sufficient to reach atomic resolution.

We used a single crystal bcc Fe film to demonstrate the use of multihole apertures for quantitative EMCD measurements. For this purpose, a 35 nm thick Fe film was epitaxially grown on a MgO (001) substrate using direct-current magnetron sputtering. The film was capped with 6 nm thick, radio-frequency sputtered MgO layer to prevent oxidation of the Fe layer. The base pressure of the growth chamber was below 3.5 x $10^{-8}$ mbar and the operating pressure of the Ar gas was 2.7 x $10^{-3}$ mbar. Prior to the deposition, the substrate was annealed for 1h at 550 °C. The substrate temperature during the growth was kept constant at 165°C. The sample was investigated in a Phillips XPert Pro MRD diffractometer (Cu K$\alpha$=1.5418 Å). X-ray reflectivity (XRR) and diffraction (XRD) measurements (not shown) confirm the layer thicknesses quoted above as well as a high crystalline quality suitable for obtaining unambiguous EMCD signals.

The TEM samples were prepared in both plan view and cross sectional geometries to carry out the experiments in a [001] and a [110] zone axis of Fe respectively. The experiments were performed on a FEI Tecnai F30 (S)TEM operating at 300 kV equipped with a post column Gatan Tridiem spectrometer. The entrance apertures of the spectrometer were replaced by the different multihole apertures described below. In order to optimize the signal/noise ratio and to reduce the influence of beam damage on the shape of the Fe-edges, a focused electron probe was scanned across the sample and an image of the CCD camera in spectroscopy mode was acquired at each scan point. The EELS spectra were extracted by taking the intensity profiles across the spectral traces on the CCD images and the spectra from the multiple scan points were integrated. The beam convergence semi-angle was set to 5 mrad for most of the experiments except for the last measurements where higher convergence semi-angles up to 10 mrad were used. The inner and outer collection angles for different apertures were adjusted to the optimum values by changing the camera length.

To test the validity of the simulations shown in Ref. [18], we designed two single signal 8-hole (SS8) apertures, also called ventilator apertures, for [001] zone axis (ZA) of bcc Fe. This ventilator aperture can acquire one component (positive or negative) of magnetic signal in one orientation and either the sample or the aperture needs to be rotated to acquire the other component of the signal. This rotation can prolong the time between the two acquisitions and getting the measurements under identical acquisition conditions (same sample area of the sample, identical orientation) might be demanding. To minimize this problem, we fixed two ventilator apertures in the spectrometer entrance aperture rotated by 22.5° with respect to each other in a way that the two apertures could acquire opposite magnetic signals for a fixed orientation of the diffraction pattern as shown in Fig. 1(a, b). The acquisition of two magnetic signals can now be accomplished by scanning a sample area twice just switching between the apertures in between the acquisitions. The TEM sample was oriented on the [001] ZA and an area of the sample was chosen for the measurements where the sample thickness was approximately 35 nm. Spectral data in the form of line scans were acquired using each of the apertures from the same region of the sample. The final EELS spectra obtained for both of the apertures were background subtracted using a power law model and the resulting spectra were post edge normalized [19]. As compared to the C+ spectrum, the C- spectrum has a weaker Fe $L_3$ edge and a stronger Fe $L_2$ edge signal as shown in Fig. 1(c). This change in edge intensity is equally shown in the difference spectrum.

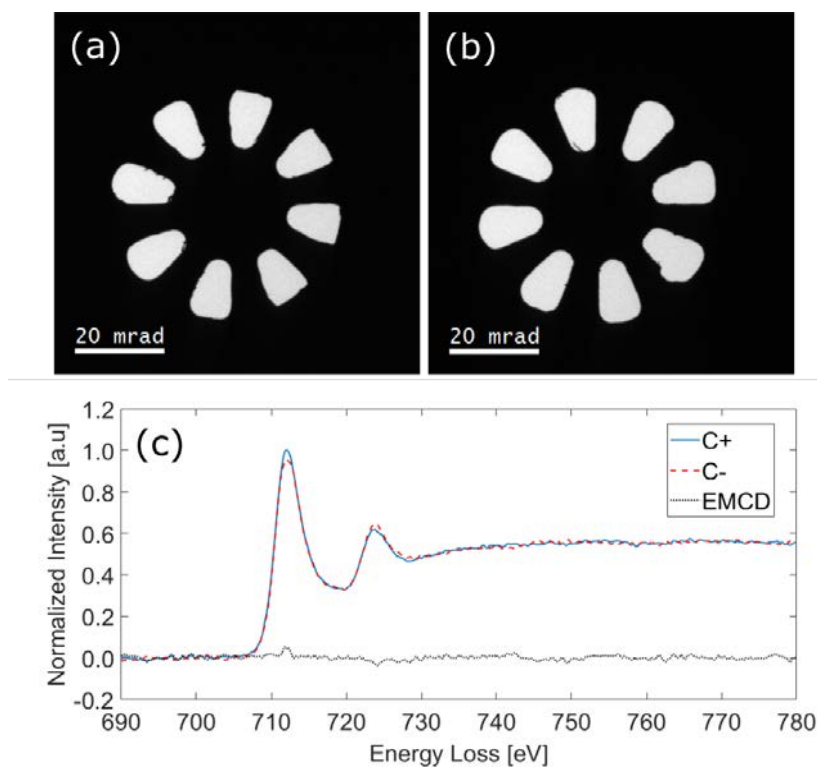

**FIG. 1.** (a) & (b) CCD images of two ventilator apertures fixed in the spectrometer entrance apertures. In STEM mode, a line scan was acquired for each aperture from the same region of the sample. The beam semi-convergence angle was 5 mrad and the inner/outer collection angles for the apertures were set to 12.5/25 mrad. A 2D image of the EELS spectrum on the CCD camera was acquired at each scan point with an acquisition time of 5s/scan point. For each line scan, an EELS spectrum was obtained by integrating the 2D data of 15 scan points and taking the intensity profile along the spectral trace. (c) The post-processed and intensity normalized spectra for the two apertures along with their difference (EMCD) signal are shown.

The ventilator apertures shown in Fig. 1 can detect the EMCD signals in ZA geometry, but their use for the acquisition of EMCD signals has a few limitations. One drawback is the complexity of the apertures themselves. The EELS spectrometer was built originally for round apertures. Here, for the ventilator apertures, in principle, each individual hole has its own spectrometer transmission function. In the alignment process of the spectrometer, it is thus very hard to align the spectrometer for all these apertures with many holes located in different directions with respect to the spectrometer axis. The resulting measurements might thus include some artifacts coming from the non-ideal spectrometer alignment. Another disadvantage of the ventilator aperture is that it can only be used for one specific crystal symmetry and a new aperture needs to be designed for every different crystal symmetry as well as for every zone axis of interest of the same crystal. Here we propose and have built a simplified form of the ventilator aperture which can be used to get EMCD measurements in multiple crystal symmetries and is much easier to align in the spectrometer. We call this aperture double signal 2-hole (DS2) aperture. The additional advantage of DS2 is that is can acquire both positive and negative magnetic signals simultaneously. In our recent work we discussed the advantages of simultaneous acquisition of the two EELS spectra for EMCD measurements [20]. We experimentally show that the DS2 aperture allows the detection of EMCD signals with beam convergence angles high enough to enable atomic resolution. To explore the behavior of DS2 aperture for different crystal symmetries, we applied it for EMCD measurements in two different zone axis [001] and [110] of bcc Fe. For the measurements on the [001] ZA, the same sample and the similar experimental conditions were used as described for the ventilator aperture. Fig. 2 shows the CCD image of the DS2 aperture, the electron diffraction pattern acquired in the region of measurement and the resulting EMCD signal. One of the concerns about the DS2 aperture could be that the signal strength or the S/N ratio would lower than the ventilator aperture but this can be compensated by increasing the dwell time/scan point or integrating more scan points in the map.

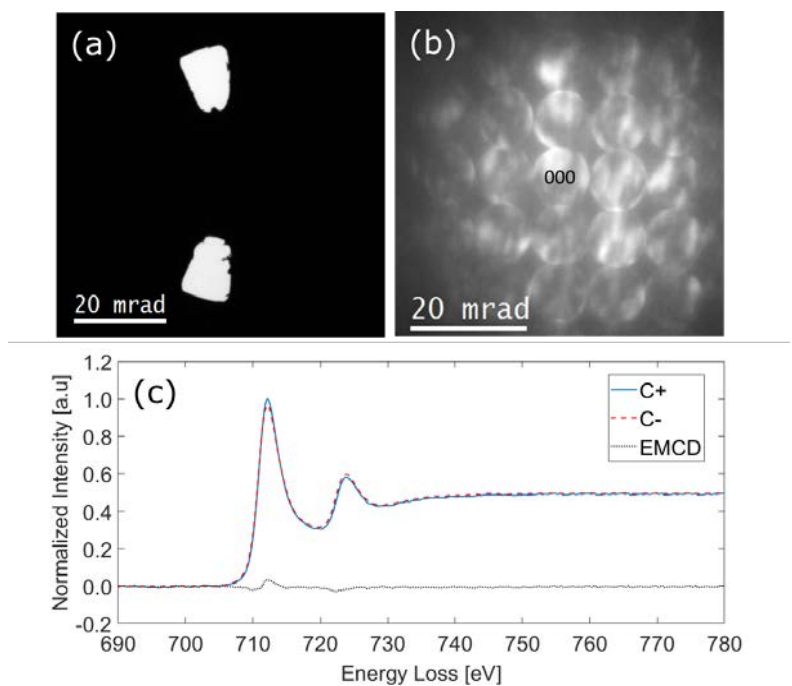

**FIG. 2.** (a) CCD image of the DS2 aperture (b) an electron diffraction pattern from the region of measurement. In the STEM mode, the electron beam with a convergence semi-angle of 5 mrad was scanned in a line across the sample and a 2D CCD image of the two EELS spectra was acquired at each scan point. The acquisition time was 5s/scan point and the inner/outer collection angles for the aperture were set to 12.5/25 mrad. Data of 60 scan points was integrated to generate a final 2D image containing two spectral traces for each aperture and the EELS spectra were extracted by taking the intensity profiles along each spectral trace. (c) The post-processed and intensity normalized spectra along with their difference are shown.

To apply the DS2 aperture in [110] ZA we have performed simulations of the inelastic electron scattering [21] of an electron beam with convergence semi-angle of 5 mrad, accelerated by 300 kV, on a 20 nm thick crystal of bcc iron. In calculations presented here, the beam direction was set along the [110] zone axis. Results are summarized in Fig. 3. Note that for zone axis [110] the symmetry of the diffraction pattern is lower, when compared to the [001] zone axis treated in Ref.[18]. Reduced symmetry of the diffraction pattern is clearly visible from both non-magnetic (Fig. 3(a)) and magnetic (Fig. 3(b)) components of the inelastic scattering cross-section at the iron $L_3$ edge. Superposing the 8-hole ventilator aperture over the magnetic signal distribution (Fig. 3(c)) shows that not all holes collect a magnetic signal of the same sign. In fact some holes collect an approximately even amount of positive and negative EMCD. This led us to design the DS2 aperture shape, which should be equally efficient for both [110] and [001] zone axes.

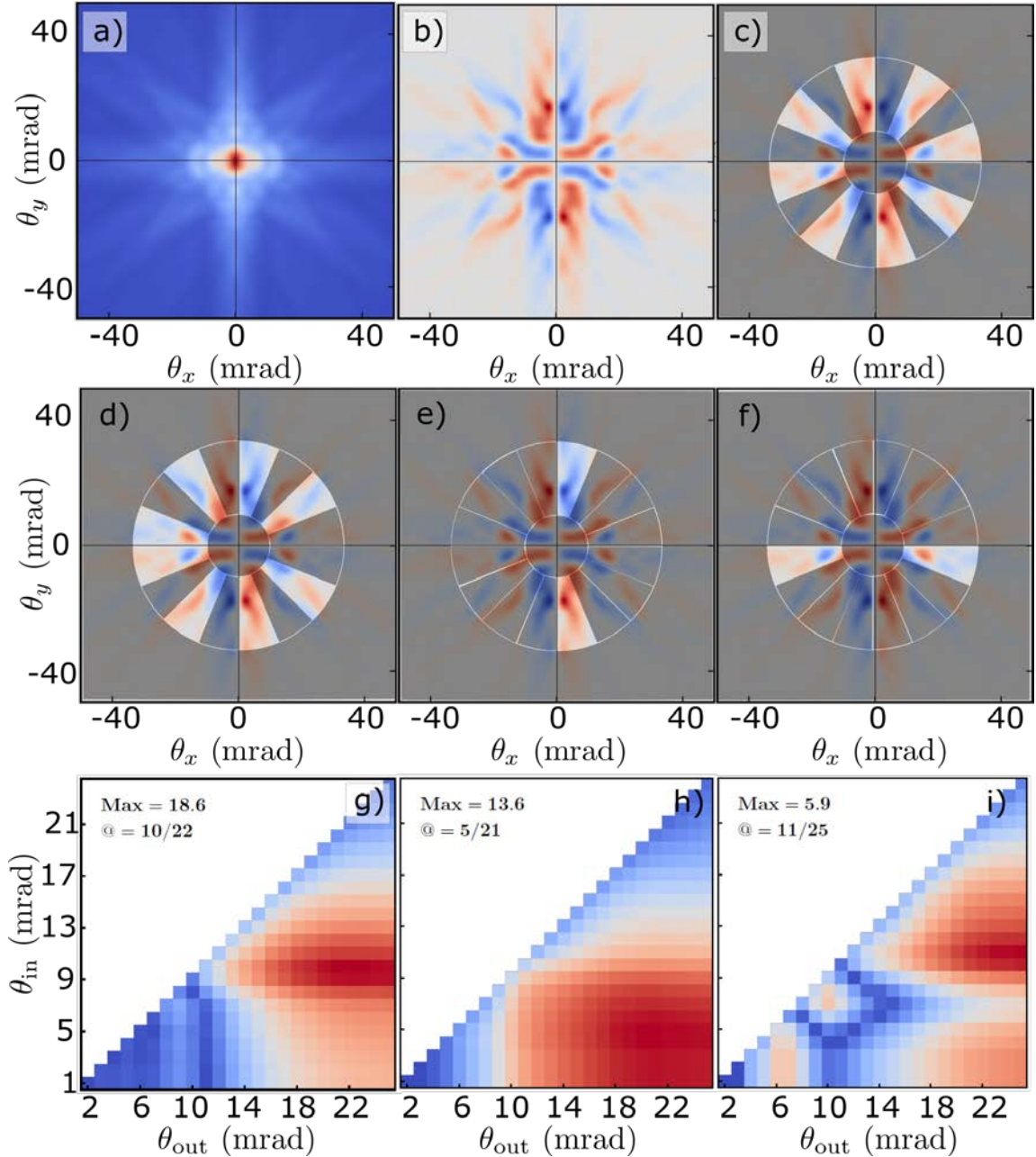

**FIG. 3.** Simulations of the non-magnetic (a) and magnetic (b) component of the inelastic scattering cross-section at the $L_3$ edge of iron. Beam of 5 mrad convergence semi-angle accelerated by 300 kV is passing parallel to [110] zone axis of 20 nm thick iron crystal. In panels (c) and (d) the ventilator and the DS7 apertures are superposed over the magnetic signal shown in panel (b). In panels (e) and (f) the DS2 aperture is superposed over the magnetic signal with aperture holes parallel to (100) and (110) axes respectively. Panel (g) shows the inner/outer collection angles calculated for ventilator aperture. Panel (h) and (i) show the optimum inner/outer collection angles calculated for the DS2 aperture orientations shown in (e) and (f) respectively.

According to the simulations for [110] ZA, the DS2 aperture can be aligned either parallel to (100) or (110) axis in the diffraction plane as shown in Fig. 3 (e, f). The latter orientation was chosen as the aperture stays further away from the high intensity first order Bragg reflections which may disturb the very weak EMCD signal. The optimum inner/outer collection angles in this case should be 11/25 mrad which are quite close to the collection angles we already used in [001] ZA. We thus used the same collection angles as in the previous experiments. A cross sectional Fe sample was used for the measurements in [110] ZA. The sample thickness was approximately 25 nm in the area of measurements. The experiment was performed for three convergence semi-angles 5 mrad, 7.5 mrad and 10 mrad of the incident electron probe. The EMCD signals obtained by taking the difference of post processed EELS spectra are shown in Fig. 4.

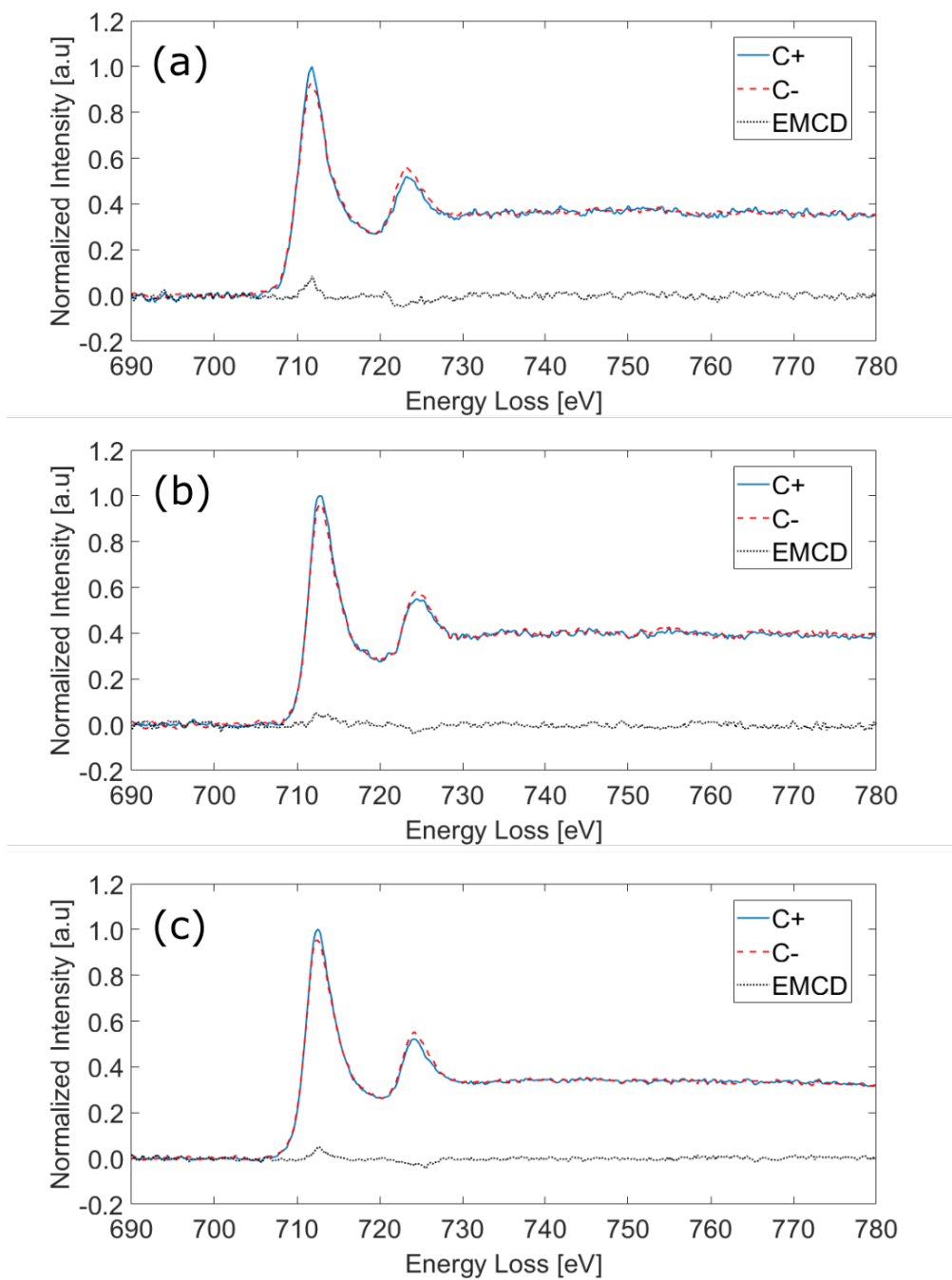

**FIG. 4.** EMCD signals acquired using DS2 aperture for beam convergence semi-angles of (a) 5 mrad (b) 7.5 mrad (c) 10 mrad. In STEM mode, the electron beam with different convergence angles was line scanned across the Fe film and a 2D CCD image of the EELS spectra was acquired at each scan point. The acquisition time for (a) and (b) was 3s/scan points and 2s/scan point for (c). The final EELS spectra were produced by integrating 30 scan points for (a), (b) and 50 scan points for (c). The inner/outer collection angle for the aperture was 12.5/25 mrad. The plus and minus EELS spectra were extracted by taking the intensity profiles along the spectral traces for the upper and lower holes of the aperture. The spectra were then background subtracted and post edge normalized.

It is worth noting that a clear EMCD signal is visible on both the $L_3$ and $L_2$ edges in all three cases given in Fig. 4. At a semi-convergence angle of 10 mrad, the diffraction limited electron probe has a diameter of 1.2 Å which is smaller than the lattice plane spacings in many magnetic materials. Thus, in an aberration corrected microscope, the set up should be capable to get EMCD measurements with atomic resolution.

In the above EMCD experiments using the ventilator apertures, a 2D CCD image of EELS spectra was acquired at each beam dwell point of a line scan instead of acquiring directly the EELS spectrum as performed in conventional STEM-EELS spectrum imaging. The acquisition scheme using the full 2D CCD image provides with a possibility to correct some of the alignment errors of the EELS spectrometer in post processing steps. It seems important to highlight this aspect since due to the complex non circular shape of the apertures, the alignment of the spectrometer is not straightforward and would need to be optimized further. We have used a MATLAB script to correct the apparent alignment errors in the spectra, see also supplemental material. Moreover, this kind of dataset gives the freedom to extract the EELS spectra from optimum scattering angles within the aperture range to minimize artifacts from high intensity diffraction spots on the EMCD signal as shown in the supplemental material. In fact we have optimized the design of double signal 7-hole (DS7) aperture proposed in Ref. [18]. We have cut aperture holes now smaller in order to avoid intensity of crystalline reflections to enter the aperture (Fig. 5(a)). For yet unknown reasons, with this aperture, the outer part of the spectral trace does not seem to be well focused. Indeed, when evaluating the inner part of the spectral trace which corresponds to the electron intensity transmitted through the biggest hole in the DS7 aperture and which now is not transmitting intensity from strong Bragg reflections, we obtain a clear EMCD signal (Fig. 5(b)). This indicates that such ventilator apertures have a better performance when the position of aperture holes is clearly distinct from the intensities of crystalline reflections.

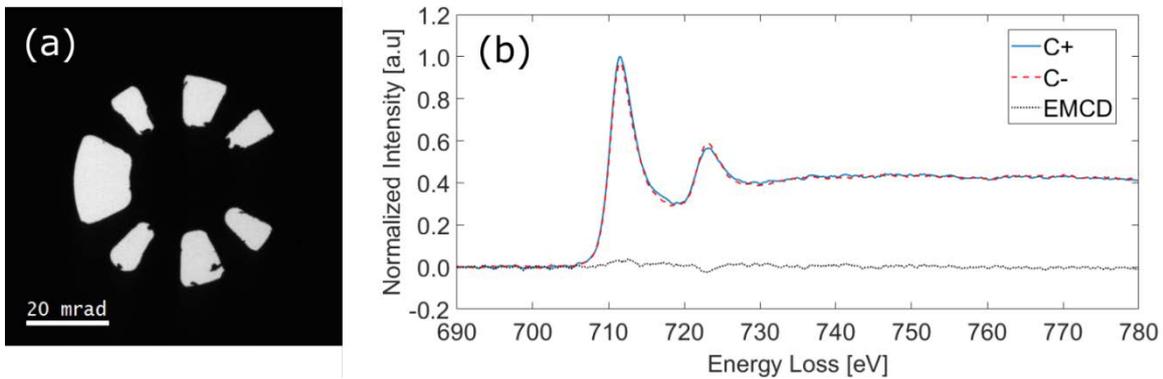

**FIG. 5.** (a) CCD image of DS7 aperture. The same number of scan points, dwell time, convergence and collection (inner/outer) angles were used as mentioned above for the ventilator apertures. The plus and minus EELS spectra were obtained by taking the intensity profiles in the upper and lower parts of the spectral trace. (b) The post-processed and intensity normalized spectra along with the EMCD signal are shown.

To check the quality and precision of the data, we have applied sum rules [22, 23] to all of the EMCD signals shown above and obtained the $m_l/m_s$ values for each of them which are shown in Table. 1. All values of orbital to spin magnetic ratio are in 0.04-0.09 range which are close to the values determined by other techniques [24, 25] and EMCD measurements [12, 26] for bcc Fe.

Table. 1. $m_l/m_s$ values calculated for EMCD signals acquired with different apertures

| Aperture type | $m_l/m_s$ values |
|---|---|
| Ventilator ([001] ZA, α=5mrad) | 0.050 ± 0.018 |
| DS2 ([001] ZA, α=5mrad) | 0.041 ± 0.010 |
| DS2 ([110] ZA, α=5mrad) | 0.045 ± 0.024 |
| DS2 ([110] ZA, α=7.5mrad) | 0.062 ± 0.020 |
| DS2 ([110] ZA, α=10mrad) | 0.045 ± 0.014 |
| DS7 ([001] ZA, α=5mrad) | 0.090 ± 0.011 |

We have used three different types of apertures to acquire EELS spectra on zone axis orientations of monocrystalline Fe in order to extract EMCD signals. The acquisition strategy has been optimized for those apertures. From all apertures, EMCD signals are obtained on both $L_3$ and $L_2$ EELS edges and obtained $m_l/m_s$ ratios are in good agreement with previous works. Using a double signal, two hole aperture to simultaneously obtain the EMCD signal from the two conjugate EELS spectra enabled to obtain an EMCD signal at a semi convergence angle of 10 mrad, which corresponds to an electron probe size of 1.2 Å. Thus, this work enables atomic scale resolved EMCD work for convergent electron beams and on a zone-axis orientation for the first time.

We acknowledge the contributions of Sarah Sanz in fabricating and optimizing the Fe films. We acknowledge Eric Lindholm for his great availability to fabricate this large variety of apertures. We gratefully acknowledge the generous support from the Swedish Science Council, VR grants numbers 2016-05259 and C0367901, the support from the Knut and Alice Wallenberg Stiftelse and the Swedish Foundation for Internationalization of Cooperation in Research and Higher Education, STINT, grant number IG2009 2017.

* Klaus.Leifer@angstrom.uu.se